      [2011/12/01 v34c Il Nuovo Cimento]
\documentclass{cimento}

\title{Doppler effect in the oscillator radiation process in the medium}
\author{Lekdar~Gevorgian\from{ins:x}
\atque
Valeri~Vardanyan\from{ins:y}
}
\instlist{\inst{ins:x} Theory Department, A. Alikhanyan National Scientific Laboratory - Alikhanyan Brothers Str. 2, 375036 Yerevan, Armenia
  \inst{ins:y} Faculty of Physics, Yerevan State University - Alex Manoogian Str. 1, 375025 Yerevan Armenia}
\PACSes{\PACSit{41.20.Jb, 41.60.Ap, 41.60.Bq}{Electromagnetic wave propagation, Radiowave propagation, Synchrotron radiation, Cherenkov radiation}
}
\begin{document}

\maketitle

\begin{abstract}
The purpose of this paper is to investigate the radiation process
of the charged particle passing through an external periodic field in a dispersive
medium. In the optical range of spectrum we will consider two cases: first, the
source has not eigenfrequency, and second, the source has eigenfrequency. In the first
case, when the Cherenkov radiation occurs, the non-zero eigenfrequency produces a
paradox for Doppler effect. It is shown that the absence of the eigenfrequency solves
the paradox known in the literature. The question whether the process is normal
(\emph{i.e.} hard photons are being radiated under the small angles) or anomalous depends
on the law of the medium dispersion. When the source has an eigenfrequency the
Doppler effects can be either normal or anomalous. In the X-ray range of the
oscillator radiation spectrum we have two photons radiated under the same angle - soft and hard. In this case the radiation obeys to so-called complicated Doppler
effect, \emph{i.e.} in the soft photon region we have anomalous Doppler effect and in the
hard photon region we have normal Doppler effect.
\end{abstract}

\section{Introduction}
In 1877 Austrian scientist Ernst Mach published a paper about supersonic apparatus' motion in a gas medium. He showed that in that case it generates a shock wave which has a sharp conical front with a semi angle $\alpha$ in the top:

\begin{equation}
\sin\alpha=u/v,
\end{equation}
$\noindent$
where $u$ is the speed of the voice in that medium and $v$ is the speed of the apparatus $(u/v<1)$.

Later, in 1888, Oliver Heaviside considered the motion of the point charge in a dielectric with a speed $v$ exceeding the speed of light $u$ in that dielectric \cite{ref:heav}. He showed that the geometrical place of the points in the fronts of spherical waves, emitted by the charge, is a conic surface and got formula (1).

In 1901 Lord Kelvin specified that the charge, which travels in the vacuum with a superluminal speed, has to create an electromagnetic wave similarly to Mach's waves in acoustics \cite{ref:kelvin}.

In 1904 Sommerfeld found that the charge which travels uniformly in the vacuum is able to emit electromagnetic waves if its speed is higher than the speed of light, though the special theory of relativity states that motion with such speed is impossible \cite{ref:somer}.

This phenomenon is known to be first observed by Pierre and Marie Curie in the bottles with the liquors of the radium salt. Experimentally the glow of the liquids irradiated by gamma-rays was investigated by Mallet \cite{ref:mallet}. Mallet found two characterizations: first, the glow is universal and second, the spectrum of glow is continuous.

To make sure that Mallet's conclusions are correct, Cherenkov according to Vavilov's advice started experimental investigation of this problem \cite{ref:cherenk}. They ascertained that this is not a luminescence and Vavilov in the work \cite{ref:vav} assumed that it is a slowing-down radiation (bremsstrahlung). Later the quantitative theory developed by Tamm and Frank \cite{ref:tammf} explained that the observed glow is much more intensive than bremsstrahlung. In 1942 Frank considered the radiation of the oscillating electrical dipole in a homogeneous dispersive medium \cite{ref:frank}. The mechanism of moving particle's radiation was clarified using the energy-momentum conservation law. Some interesting details are summed up in the work \cite{ref:ginz}.

In this work we will investigate the Doppler effect in the oscillator radiation process. We will clear out when the effect is normal or anomalous and when we deal with complicated Doppler effect. The complicated Doppler effect takes place in the relativistic particle's radiation process when the medium has an eigenfrequency (for example in a resonant transition radiation \cite{ref:Ter} or in a crystal \cite{ref:nitta}) and when the particle has an eigenfrequency \cite{ref:baryshevsky}. Under the given angle there are two emitted photons with different frequencies and the difference of these frequencies depends on the particle energy.

\section{Electrodynamics, conservation laws and Doppler effect}

Let the electron, with charge $e$ and with speed $v$ in the longitudinal direction $z$, oscillate in the transversal direction $x$ with frequency $\Omega$ in a homogeneous dispersive medium for which the Fourier transform of dielectric permittivity is $\epsilon(\omega)$. For the Fourier transform of the electron radiation field at the distance R (in the wave zone) we have \cite{ref:jackson}:

\begin{eqnarray}
& &
E(\omega)=\frac{e\omega}{\sqrt{2\pi}Rc}\int[\vec{n}\times(\vec{n}\times\vec{\beta}\sqrt{\epsilon(\omega)})]e^{i\omega(1-\beta\sqrt{\epsilon(\omega)}\cos\theta)t-i\alpha_p \sin\Omega t}dt, \\
& &
\alpha_p=\beta_\perp\frac{\omega}{\Omega}\beta\sqrt{\epsilon(\omega)}\sin\theta_p\cos\varphi,\nonumber \end{eqnarray}
$\noindent$
where $\vec{n}=\vec{R}/R$ is the unit vector, directed from the point of the charge location to the observation point, $\theta$ is the vectorial angle.

We can find the frequency-angular distribution of the $p$-th harmonic's radiation intensity with unit length of path using the following formula \cite{ref:gev}:

\newpage
\begin{eqnarray}
& & \frac{dW_p}{d\omega dOdz} =
\frac{e^2\omega\beta\sqrt{\epsilon(\omega)}}{2\pi c^2}[\sin^2\theta_p-p\frac{\beta_\perp}{\beta\alpha_p}\sin^2\theta_p\cos\varphi \\
& & +p^2\frac{\beta^2_\perp}{\beta^2\alpha^2_p}(1-\sin^2\theta_p\cos^2\varphi)]
J^2_p(\alpha_p)\frac{\sin^2z_p}{z^2_p},\nonumber
\end{eqnarray}
$\noindent$
where $\omega$ is the radiation frequency, $dO=sin\theta_p d\theta_p d\varphi$ is the solid angle, $\theta_p$ is the emission angle of the $p$-th harmonic, $\varphi$ is the azimuth angle,  $\beta=v/c$, $c$ being the speed of light in the vacuum, $\beta_\perp$ is the maximum angle of electron's deflection from the $z-$axis, $J_p$ the Bessel function. Here $\alpha_p$ and $z_p$ are determined by the following expressions:

\begin{equation}
\alpha_p=\frac{\omega}{\Omega}\beta_\perp\sqrt{\epsilon(\omega)}\sin\theta_p\cos\varphi,
z_p=\frac{\omega}{\Omega}\pi n(1-\beta\sqrt{\epsilon(\omega)}\cos\theta_p-\frac{p\Omega}{\omega}),
\end{equation}
$\noindent$
where $n$ is the number of oscillation periods. The last multiplier is the factor of periodicity of electron's motion, which with an accuracy of $1/n$ becomes the Dirac delta-function:

\begin{equation}
\lim_{n \to \infty}\frac{\sin^2z_p}{z^2_p}=\frac{\Omega}{n\omega}\delta(1-\beta\sqrt{\epsilon(\omega)}\cos\theta_p-\frac{p\Omega}{\omega}).
\end{equation}

Here $p$ can be any integer number which satisfies the delta-function definition.

Thus Maxwell equations' solution for the problem of the oscillator radiation in the medium includes the energy-momentum conservation law. It is reflected in the argument of the delta-function. The phase of the Fourier transform of the radiation field does not depend on time. The stationary phase of the radiated wave follows from Huygens' principle. On the other hand, we can find radiated frequency $\omega$ dependence on the source frequency $\Omega$ and radiation angle $\theta_p$ from expression (6):

\begin{equation}
\omega=\frac{p\Omega}{1-\beta\sqrt{\epsilon(\omega)}\cos\theta_p}.
\end{equation}

When the source is moving we detect another frequency of radiation than if the source was in the rest. This phenomenon is known as Doppler effect.

Note that in the vacuum ($\epsilon(\omega)=1$), when $p\geq1$ the hard photons are being radiated under small angles. It is called a normal Doppler effect. In the opposite case the effect is anomalous.

\section{Doppler effect in the optical range of spectrum}

\subsection{Source without eigenfrequency}

In this case we can deal with Cherenkov radiation ($\beta\sqrt{\epsilon(\omega)}>1$). One can see that for the Cherenkov radiation the frequency $\omega$ does not become infinite, as noted in \cite{ref:ginz}, but is determined by the condition $1-\beta\sqrt{\epsilon(\omega)}\cos\theta_p=0$. As $p=0$, the formula (6) goes to the 0/0 uncertainty. Note that the effect can be either normal (higher the frequency and smaller the angle) or anomalous (higher the frequency and higher the angle). It depends on the behavior of the $\epsilon(\omega)$ function in the dispersion region. The shape of that function depends on the characteristics of the medium. In the medium where the polarization occurs due to electron-ion mechanism (diamond, oil, etc.) the $\epsilon(\omega)$ is an increasing function (the dispersion is positive). In the case when the polarization occurs due to the orientation mechanism (water) the $\epsilon(\omega)$ is a decreasing function (the dispersion is negative). Note that from the $\cos\theta_p(\omega)=1/\beta\sqrt{\epsilon(\omega)}$ it is easy to see that in the first case the effect is anomalous and in the second case the effect is normal.

From expression (3) we obtain the formula for the intensity of Cherenkov radiation ($p=0$). The source does not have eigenfrequency and the relationship between frequency and angle of radiation is determined by equating the argument of the delta-function to zero. That relationship is defined by the energy-momentum conservation law and not by the Doppler effect.

\subsection{Source with eigenfrequency}

Relationship (6) between the radiation angle and frequency is useful to write the following:

\begin{equation}
\cos\theta_p=\frac{1-p\Omega/\omega}{\beta\sqrt{\epsilon(\omega)}}.
\end{equation}

For the optical range of spectrum when the Cherenkov radiation does not occur ($\beta\sqrt{\epsilon(\omega)} \leq 1$) we can mention that $p \geq 1$. When we have Cherenkov radiation ($\beta\sqrt{\epsilon(\omega)} > 1$), $p$ can be both negative and positive.

When $p$ is positive and the medium has a negative dispersion, the Doppler effect is normal (\emph{i.e.} when the frequency increases, then the angle decreases). If the medium dispersion is positive, the Doppler effect can be either normal or anomalous: when the $\sqrt{\epsilon(\omega)}$ increases slower than the numerator ($1-p\Omega/\omega$), then the effect is normal and in the opposite case the effect is anomalous.

For negative $p$ ($1\leq|p|<(\beta\sqrt{\epsilon(\omega)}-1)\omega/\Omega$), if the medium has a positive dispersion, the Doppler effect is anomalous. When the dispersion of the medium is negative the Doppler effect can be either normal or anomalous: if the $\sqrt{\epsilon(\omega)}$ decreases faster than the numerator ($1-p\Omega/\omega$), then the effect is normal and in the opposite case the effect is anomalous.

One can mention that as noted in the work \cite{ref:gev} in practical conditions the optical undulator radiation is negligible compared with the Cherenkov radiation.

Note that the positivity of the frequency is ensured not by putting the module sign in the denominator \cite{ref:ginz}, but by the fact that for the negative sign of the denominator the emitted harmonics have negative numbers.

\section{Doppler effect in the X-ray range of spectrum}

In this case $\omega$ is much larger than the plasma frequency of medium $\omega_p$ and the dielectric permittivity's dispersion is expressed as follows:

\begin{equation}
\epsilon(\omega)=1-\frac{\omega^2_p}{\omega^2}.
\end{equation}

Substituting (8) into (7) and taking into account that $\cos\theta_p\leq1$, we can find that $\omega$ is in the following interval:

\begin{equation}
\omega_1=\frac{p\Omega}{2(1-\beta)}(1-\sqrt{1-\frac{2(1-\beta)\omega^2_p}{(p\Omega)^2}})\leq\omega\leq\frac{p\Omega}{2(1-\beta)}(1+\sqrt{1-\frac{2(1-\beta)\omega^2_p}{(p\Omega)^2}})=\omega_2.
\end{equation}

Here, the energy of the emitted photon has lower and upper boundaries and the energy of the particle has a lower threshold.

The relationship between the radiation angle and the frequency is the following:

\begin{equation}
\cos\theta_p(\omega)=1-\frac{(1-\beta)}{\beta}(1-\frac{\omega_1}{\omega}){(\frac{\omega_2}{\omega}-1}).
\end{equation}

Since the last relationship is quadratic, there are two photons radiated under the given angle: soft and hard (under the zero angle we have radiated photons with $\omega_1$ and $\omega_2$). One can be sure that $\cos\theta_p(\omega)$ has a minimum for the frequency  $\omega_c=\omega^2_p/p\Omega$:

\begin{equation}
\cos\theta_p(\omega_c)=1-\frac{(1-\beta)}{\beta}\frac{(\omega_1-\omega_2)^2}{4\omega_1\omega_2}.
\end{equation}

So the angel $\theta_p(\omega)$ has a maximum (for $\omega=\omega_c$). The function $\cos\theta_p(\omega)$ in the interval from $\omega_1$ to $\omega_c$ decreases and $\theta_p(\omega)$ increases (the Doppler effect in this interval is anomalous). In the interval from $\omega_c$ to $\omega_2$ the effect is normal. Therefore the Doppler effect is complicated.

\section{Conclusion}

The formula for frequency-angular distribution of the intensity of oscillator's radiation in the dispersive medium contains a factor of periodicity. For large values of periods this factor becomes a delta-function. In the argument of that delta-function the energy-momentum conservation law is reflected, which is to say that the phase of Fourier transform of the radiation field is stationary. This stationarity of the phase follows from the Huygens principle.

Zeroth-order harmonic corresponds to the radiation of the source without eigenfrequency, \emph{i.e.} to the Vavilov-Cherenkov radiation. The conservation law gives a relationship between the frequency and angle of radiation and, depending on the medium dispersion law, it can be either normal or anomalous. The formula of the radiation intensity in this case coincides with the Tamm and Frank formula, and the oscillator radiation can be described by both positive and negative harmonics. If the Cherenkov radiation condition is not satisfied, the radiation of oscillator can be described only by positive harmonics.

Non-zero harmonics are the radiation of the source with an eigenfrequency and then we deal with the Doppler effect. In the optical range of spectrum the Doppler effect can be either normal or anomalous. In the X-ray range one has to deal with a complicated Doppler effect: it is anomalous in the soft frequency region and normal in the hard frequency region.

\end{document}